\documentclass[twocolumn,english,aps,prl,reprint,superscriptaddress,amsmath,amssymb,floatfix]{revtex4-1}
\usepackage[T1]{fontenc}
\usepackage[latin9]{inputenc}
\usepackage{xcolor}
\usepackage{amsmath}
\usepackage{stmaryrd}
\usepackage{graphicx}
\PassOptionsToPackage{normalem}{ulem}
\usepackage{ulem}

\usepackage{color}

\makeatletter

\usepackage{babel}

\makeatother

\usepackage{textcomp}

\begin{document}
\title{First-order transition between the plaquette valence bond solid and antiferromagnetic phases of the
Shastry-Sutherland model}
\author{Ning Xi}
\thanks{These authors contributed equally to this study.}
\affiliation{Department of Physics and Beijing Key Laboratory of Opto-electronic
Functional Materials and Micro-nano Devices, Renmin University of
China, Beijing 100872, China }
\author{Hongyu Chen}
\thanks{These authors contributed equally to this study.}
\affiliation{Department of Physics and Beijing Key Laboratory of Opto-electronic
Functional Materials and Micro-nano Devices, Renmin University of
China, Beijing 100872, China }
\author{Z. Y. Xie}
\email{qingtaoxie@ruc.edu.cn}

\affiliation{Department of Physics and Beijing Key Laboratory of Opto-electronic
Functional Materials and Micro-nano Devices, Renmin University of
China, Beijing 100872, China }
\author{Rong Yu}
\email{rong.yu@ruc.edu.cn}

\affiliation{Department of Physics and Beijing Key Laboratory of Opto-electronic
Functional Materials and Micro-nano Devices, Renmin University of
China, Beijing 100872, China }
\begin{abstract}
We study the ground state phase diagram of the Shastry-Sutherland
model by using the variational optimization of the infinite tensor network
states, and find a weakly first-order transition between the plaquette
and the antiferromagnetic states. The
full plaquette state strongly competes with the empty plaquette ground
state, with an energy difference less than $10^{-4}J$. We show a staggered
ring exchange interaction that preserves the Shastry-Sutherland lattice
symmetry can stabilize the full plaquette ground state. In light of this, we propose
the triple point where the full plaquette, empty plaquette, and antiferromagnetic
phases meet as a deconfined quantum critical point.
\end{abstract}
\maketitle

\textit{Introduction.} Enhanced quantum fluctuations in frustrated
spin system can give rise to exotic quantum phases, including quantum spin liquid (QSL), valence bond solid (VBS), and spin nematicity \citep{Diep_Book, Mila_Book, Balents_Nature:2010, Savary_2016, ZhouYi2017, Yu_PRL:2015}.
The nature of these novel quantum phases and related quantum phase
transitions have been extensively studied. Though most transitions
can be described within the standard Ginzburg-Landau-Wilson (GLW)
paradigm, it has been proposed that the transition between a VBS and
an antiferromagnetic (AFM) phase is beyond the GLW scenario, \emph{e.g.},
via a deconfined quantum critical point (DQCP) \citep{Senthil2004}.
At this point, deconfined fractionalized excitations emerge, and the
enhanced symmetry allows a continuous rotation between the two distinct
order parameters. However, besides some sophisticatedly designed models~\citep{Sandvik2007, Meng_EasyplaneDQCP, Shao_Science},
it is still challenging to realize a DQCP in two-dimensional (2D)
frustrated spin systems. 

\begin{figure}
\includegraphics[width=82mm]{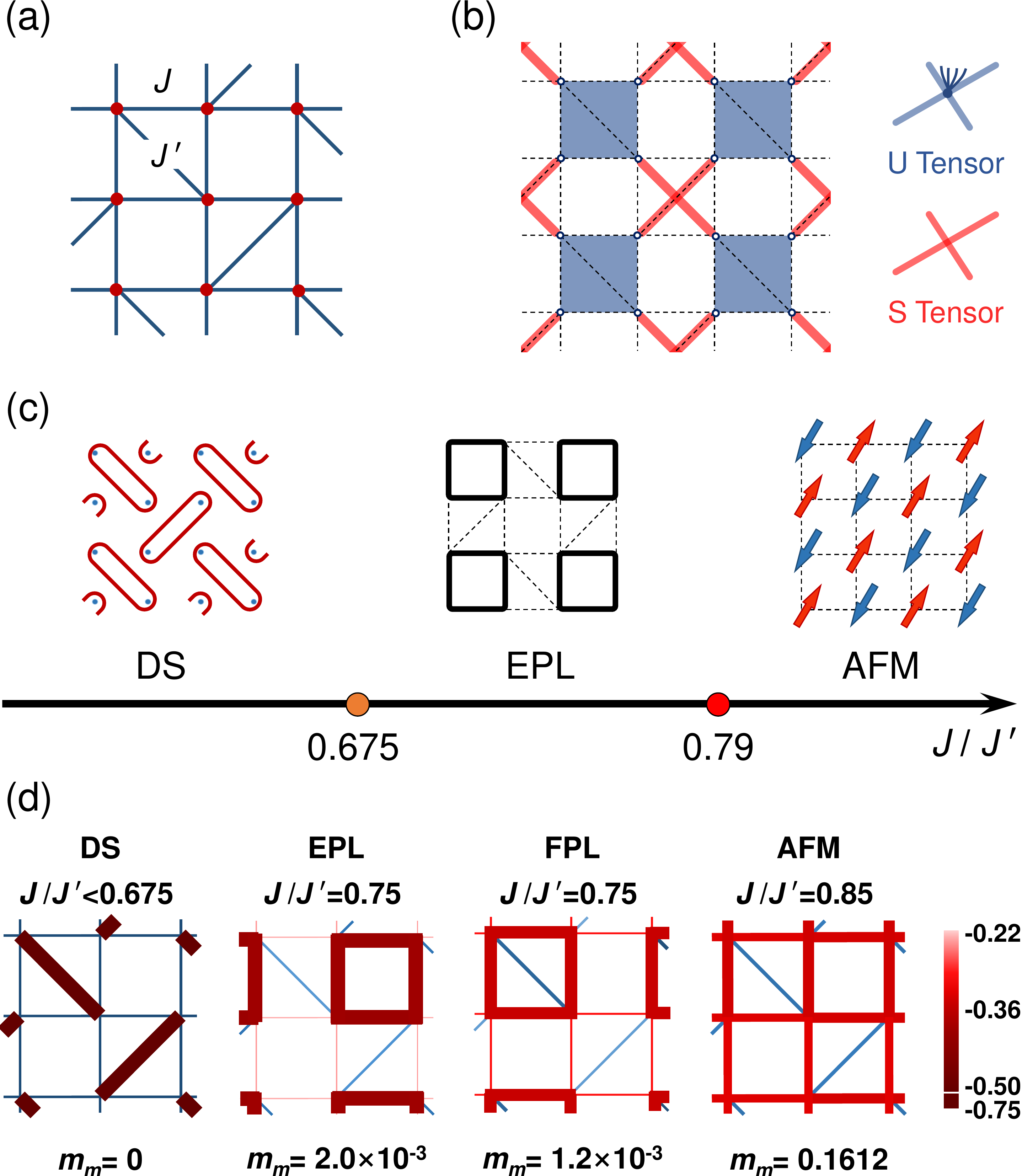} \caption{(a) Sketch of the SS model. $J$ and $J^{\prime}$ refers to the nearest-
and next-nearest-neighbor Heisenberg couplings. (b) The 16-PESS setup
on the SS lattice. The tensors are placed at the plaquette within
the $J^\prime$ coupling. Each $U$ tensor (blue) consists of $4$ physical
spin indices and $4$ auxiliary ones, while an $S$ tensor contains
$4$ auxiliary indices only.
(c) The ground-state phase diagram of
the SS model. DS,
EPL, and AFM refer to dimer singlet, empty plaquette, and antifferromagnetic
phases, respectively. Both the DS-EPL and EPL-AFM transitions are found to be first-order.
(d) Typical configurations, corresponding spin-spin correlations and magnetic order parameters of the DS, EPL, FPL, and AFM states in the calculation. The EPL and FPL have distinct characters
of symmetry breaking. \label{fig:1}}
\end{figure}

The Shastry-Sutherland (SS) model~\citep{Shastry1981} is an ideal
frustrated spin model for studying the VBS-AFM transition~\citep{Koga2000, Chung2001, Pixley2014, Corboz2013, Boos2019, Lee2019, yang2021quantum, Sengupta_PRL:2013}.
It is defined on the SS lattice as sketched in Fig~\ref{fig:1}(a),
and the Hamiltonian reads
\begin{equation}
\hat{H}_{{\rm {SS}}}=J\sum_{\langle i,j\rangle}\mathbf{S}_{i}\cdot\mathbf{S}_{j}+J^{\prime}\sum_{\langle\langle i,j\rangle\rangle}\mathbf{S}_{i}\cdot\mathbf{S}_{j},\label{eq:SS_Ham}
\end{equation}
where $\mathbf{S}_{i}$ is an $S=1/2$ spin on site $i$, $J$ and
$J'$ refer to the nearest and next-nearest neighbor couplings, respectively.
As demonstrated in Fig.~\ref{fig:1}(c), the ground state is found
to be a product of dimer singlets (DSs) \citep{Koga2000,Corboz2013} along the diagonal directions
for $J/J^{\prime}\lesssim0.68$. For $J/J^{\prime}\gtrsim0.68$,
the ground state first changes to a plaquette VBS, then to a
N\'{e}el AFM with increasing $J/J^{\prime}$ \citep{Koga2000,Corboz2013,Lee2019}.
A first-order transition between the DS and plaquette phases has been verified
by various numerical results. However, the understanding of the nature of the plaquette-AFM
transition remains controversial: A series expansion study~\citep{Koga2000}
found a second-order transition, while an infinite projected entangled
pair state (iPEPS) tensor network calculation~\citep{Corboz2013}
showed it to be weakly first-order. A recent DMRG study\citep{Lee2019}
proposed the transition is through a DQCP with an emergent $O(4)$
symmetry, but another DMRG work\citep{yang2021quantum} suggested
a gapless QSL settles in between the plaquette and AFM phases.

The SS model is believed to properly describe the quantum magnetism
of the quasi-2D material $\mathrm{SrCu}_{2}(\mathrm{BO}_{3})_{2}$ \citep{Kageyama1999,PhysRevLett.82.3701,Koga2000}.
Evidence of evolution from the DS to a plaquette then to an AFM state under pressure has
been cumulated via inelastic neutron scattering (INS)~\citep{Zayed2017},
nuclear magnetic resonance (NMR)~\citep{Waki2007,haravifard2016crystallization},
Raman scattering~\citep{Bettler2020}, and specific heat~\citep{Guo2020,Jimenez2021}
measurements. Some experimental results imply the intermediate plaquette ground state
has a full plaquette (FPL) pattern, that is, the local singlet
spans on the plaquette with a diagonal $J_{2}$ bond~\citep{Zayed2017,Bettler2020, Cui2021}.
However, numerical calculations on the SS model~\citep{Corboz2013,Boos2019,Lee2019,yang2021quantum}
suggest an empty plaquette (EPL) ground state (Fig.~\ref{fig:1}(c)).
Though it was shown theoretically that the FPL can be stabilized as
the ground state when the SS lattice symmetry is broken~\citep{Boos2019}, the suggested symmetry breaking has
not been observed. Therefore, it is still an open question how to
reconcile the discrepancy between theory and experiments.

In this Letter, we investigate the ground-state phase diagram of the
SS model by using a variational optimization of the infinite tensor
network states with the projected entangled simplex state (PESS) construction.
Our calculation clarifies the intermediate phase to be an EPL, and our results evidence a weakly first-order EPL-AFM transition
at $J/J^{\prime}\approx0.79$, without a spin liquid
phase in between. Nevertheless, the FPL state is found to be intimately competing with the EPL ground state, with an energy difference less than $10^{-4}J$. By including a staggered ring exchange interaction in the
Hamiltonian, we show that with a perturbation that preserves
the SS lattice symmetry, the ground state changes from the EPL to
FPL. In light of this observation, we construct a global phase diagram
of this generalized SS model, and propose the triple point among the
full plaquette, empty plaquette, and antiferromagnetic phases as a
deconfined quantum critical point. 

\textit{Model and method.} We consider the ground states of the SS
model defined in Eq.~\eqref{eq:SS_Ham}. To investigate the stability
of the EPL versus FPL, we also consider a generalized SS model by
including a staggered ring-exchange interaction $\hat{H}_{{\rm {Q}}}$
between the nearest neighbor spin pairs of a plaquette with $J^\prime$ interaction (see inset of Fig.~\ref{fig:4}(a)):
\begin{equation}
\hat{H}_{{\rm {Q}}}=-Q\sum_{ijkl\in\boxslash}\left(\mathbf{S}_{i}\cdot\mathbf{S}_{j}\right)\left(\mathbf{S}_{k}\cdot\mathbf{S}_{l}\right)+\left(\mathbf{S}_{i}\cdot\mathbf{S}_{k}\right)\left(\mathbf{S}_{j}\cdot\mathbf{S}_{l}\right).\label{eq:Q_Ham}
\end{equation}

The ground states are obtained by using variational
optimization of the 16-PESS tensor network states~\cite{SM}. The PESS construction
of the tensor network states has been shown to give an excellent description
to the ground state of frustrated spin systems~\citep{Xie2014,Liao2017}.
On the SS lattice, the 16-PESS is constructed with $U$ and $S$ tensors
as illustrated in Fig.~\ref{fig:1}(b). A projection tensor $U$ is defined
on a plaquette with the $J^\prime$ interaction, and carries the four spins
in this plaquette. Four adjacent $U$ tensors are then connected
by an entangled simplex tensor $S$. The $S$ tensor is introduced
to describe the entanglement among the spin clusters but itself
does not carry any physical spin degree of freedom. In this work, we find a $2\times2$ unit cell with one independent
pair of $U$ and $S$ tensors is sufficient to characterize the ground state. The calculation is performed in
an infinitely large system by employing the translational symmetry.

To determine the ground states, we adopt an advanced variational optimization
method to globally minimize the ground-state energy ${\langle\psi|\hat{H}|\psi\rangle}/{\langle\psi|\psi\rangle}$.
We are inspired by differentiable programming~\citep{Liao2019, ADSRG2020, ADPEPS2021}, which can be effectively combined with other well-developed techniques \citep{Jiang2008, Corboz2014, Phien2015, Corboz2016, wang2011cluster}.
To be specific, the state for optimization is initialized from an arbitrary state or an approximately converged state obtained from either simple update \citep{Jiang2008, Xie2014} or cluster update \citep{wang2011cluster}. Then we use the corner transfer matrix renormalization
group (CTMRG) method \citep{Corboz2014} to contract the infinite network
and get the approximate environment of the local tensors. After that,
we use the quasi-Newton L-BFGS algorithm to minimize the
energy density, 
which can be effectively implemented by the Zygote package \citep{Zygote}. The automatic differentiation provides a global optimization strategy of the ground state, and is probably more reliable than local optimization apporaches, especially for critical systems where many competing states exist.


\begin{figure*}
\centering \includegraphics[width=0.9\textwidth]{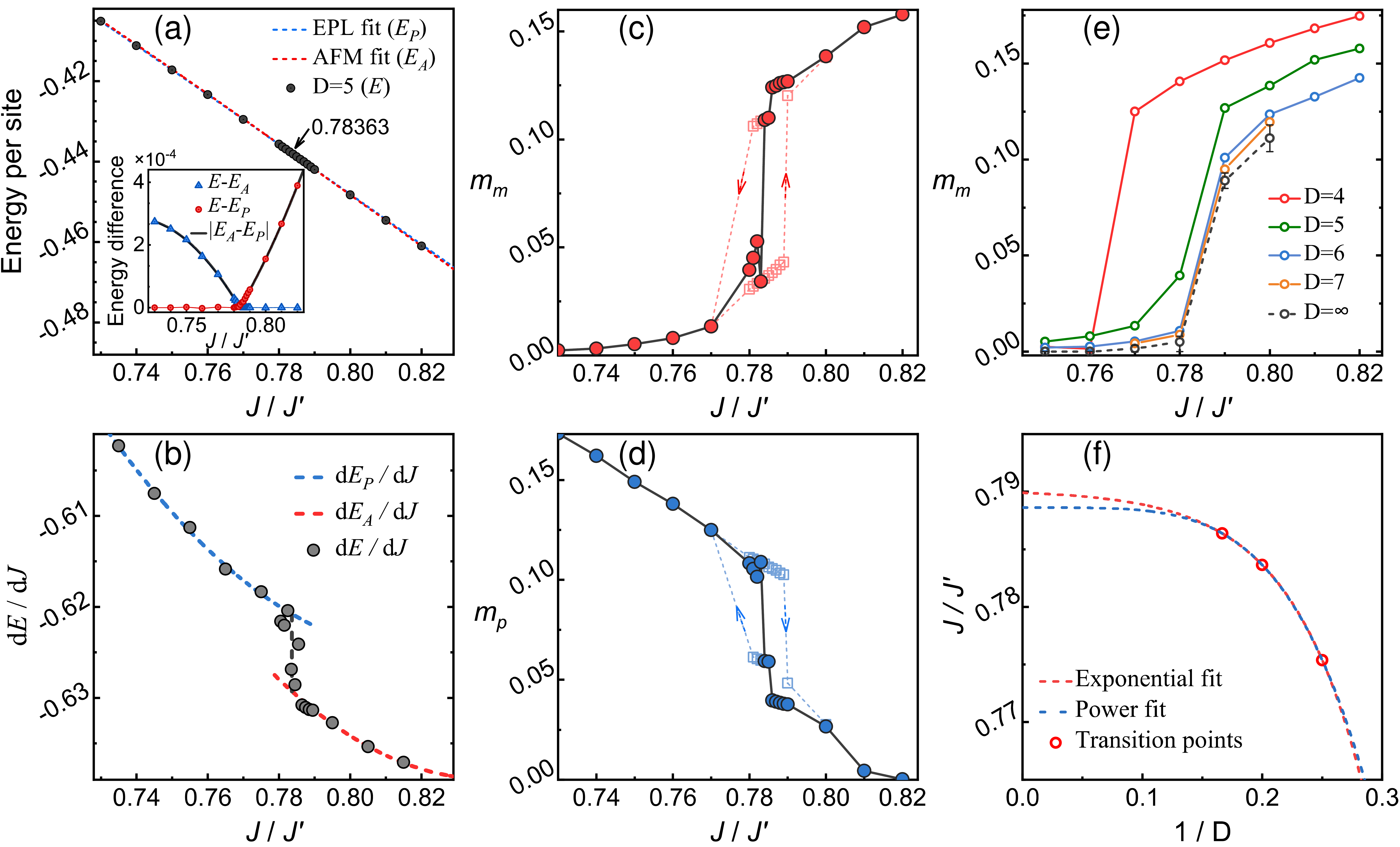} \caption{(a) The ground-state energy per site with $J/J^{\prime}$ at $D=5$.
Blue (red) dashed line is the fitted energy $E_{A}$ ($E_{P}$) in
the AFM (Plaquette) phase (see text). The insert shows the energy differences
$E-E_{A}$ and $E-E_{P}$. (b) The first-order derivative $dE/dJ$
with $J/J^{\prime}$ shows a clear discontinuity at the plaquette-AFM
transition. (c) The staggered magnetization $m_{m}$ with
$J/J^{\prime}$ for $D=5$. Closed circles show the $m_{m}$ value of lowest
energy configurations, while the open squares are obtained with biased configurations (see maintext) and exhibit clear hysteresis
loop. (d) Same to (c) but for the plaquette order parameter $m_{p}$.
(e) Finite-$D$ analysis and the extrapolated $m_{m}$ with $J/J^{\prime}$.
(f) Extrapolation of the transition points with $1/D$. The transition
point at each finite $D$ is determined from the crossing point of
the $E-E_{A}$ and $E-E_{P}$ curves in the inset of panel (a). \label{fig:2}}
\end{figure*}

\textit{First-order plaquette-AFM transition.} As demonstrated in Fig.~\ref{fig:1}(c),
we find the ground state is the DS for $J/J^{\prime}<0.675$, consistent
with previous works~\citep{Lee2019,Corboz2013,Koga2000}. Increasing
$J/J^{\prime}$ the ground state first changes to the EPL, then to
the AFM at $J/J^{\prime}\approx0.79$. To examine the plaquette-AFM transition,
we calculate the ground-state energy $E$ and its derivative $dE/dJ$.
The results for $D=5$ are shown in Fig.~\ref{fig:2}(a) and (b),
respectively. Though $E$ varies smoothly across the transition, $dE/dJ$
shows a small discontinuity at $J/J^{\prime}\approx0.78$, featuring 
a weakly first-order transition. We further calculated the order parameters
$m_{m}$ and $m_{p}$ of the AFM and plaquette phases, respectively.

\begin{align}
m_{m} & =\sqrt{\langle m_{m}^{x}\rangle^{2}+\langle m_{m}^{y}\rangle^{2}+\langle m_{m}^{z}\rangle^{2}},\label{eq:mm}\\
m_{p} & =\bigg|\sum\limits _{\left\langle ij\right\rangle \in\boxempty_{A}}\langle\mathbf{S}_{i}\cdot\mathbf{S}_{j}\rangle-\sum\limits _{\left\langle ij\right\rangle \in\boxempty_{B}}\langle\mathbf{S}_{i}\cdot\mathbf{S}_{j}\rangle\bigg|.\label{eq:mp}
\end{align}

Here $\left\langle m_{m}^{\alpha}\right\rangle =\frac{1}{N}\sum_{i} \left\langle e^{i\mathbf{Q}\cdot\mathbf{r}_{i}}S_{r_{i}}^{\alpha}\right\rangle _{r_{i}}$ for $\alpha=x,y,z$ and $\mathbf{Q}=(\pi,\pi)$, and $\boxempty_{A/B}$ 
label the two inequivalent empty plaquettes. As
shown by the solid curves with closed symbols in Fig.~\ref{fig:2}(c)
and (d), $m_{m}$ and $m_{p}$ both exhibit small but finite abrupt jumps
near $J/J^{\prime}\approx0.78$, consistent with the $dE/dJ$ result. To further verify the first-order
nature of the transition, we first stabilize the plaquette (AFM) state from
our optimization at a $J/J^{\prime}$ ratio far from the transition,
then slowly increase (decrease) $J/J^{\prime}$. At each step, we
take the converged state obtained from last step as the initial state for optimization. We repeat this procedure until the system is driven through the transition to the AFM (plaquette)
phase. As shown by the dashed curves with open symbols in Fig.~\ref{fig:2}(c)
and (d), both $m_{m}$ and $m_{p}$ exhibit clear hysteresis loops,
indicating the existence of metastable states, which is a prominent
signature of a first-order transition.

This first-order transition is shown in Fig.~\ref{fig:2}(e) for all the finite $D$
values we studied, and the large-$D$ limit is obtained by extrapolation. It shows clearly that in each curve the AFM order parameter $m_{m}$ experiences a jump. Though the discontinuity $\Delta m_{m}$
at the jump reduces with increasing $D$, it remains finite in the large-$D$ limit.
This evidences a weakly first-order transition in the thermodynamic
limit.

The transition point can alternatively be determined from the calculated energies
at finite $D$. As shown in Fig.~\ref{fig:2}(a), for each $D$ we
fit the energies away from the transition ($J/J^{\prime}<0.78$
in the plaquette phase and $J/J^{\prime}>0.79$ in the AFM phase
for $D=5$) to polynomial functions~\cite{SM}. Denote $E_{P}$ and $E_{A}$
to be the fitted energies in the plaquette and AFM phases, respectively,
and define $\delta E_{P(A)}=E-E_{P(A)}$. The transition point can
then be accurately determined as the crossing point of $\delta E_{P}$
and $\delta E_{A}$, as demonstrated in the inset of Fig.~\ref{fig:2}(a).
The transition points at finite $D$ values determined in this way
are plotted in Fig.~\ref{fig:2}(f). The estimated transition points in the large-$D$ limit from two different fits are both converged to $(J/J^{\prime})_{c}\approx0.79$, which is also consistent with the value obtained from $m_{m}$.

\begin{figure}
\includegraphics[width=1\columnwidth]{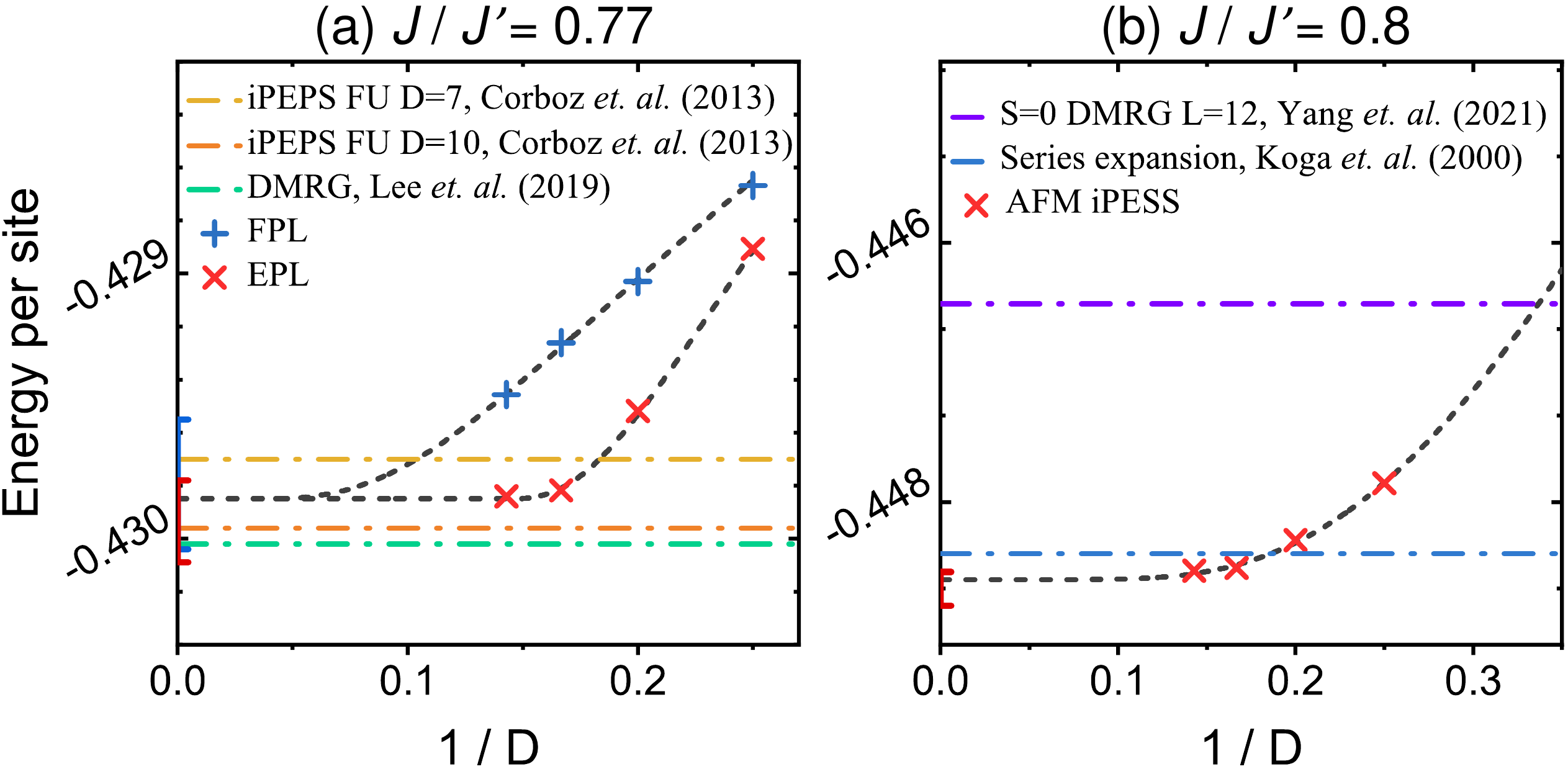} \caption{(a) Energies of EPL and FPL states versus $1/D$ at $J/J^{\prime}=0.77$.
The horizontal dashed dot lines show the plaquette ground-state energies
in an iPEPS~\citep{Corboz2013} and a DMRG~\citep{Lee2019} study
for comparison. (b) The AFM ground-state energy versus $1/D$ at $J/J^{\prime}=0.8$.
The horizontal dashed dot lines show the AFM ground-state energies
from another DMRG~\citep{yang2021quantum} and a series expansion~\citep{Koga2000}
calculation for comparison.\label{fig:3}}
\end{figure}

\begin{figure}[t]
\centering \includegraphics[width=0.98\columnwidth]{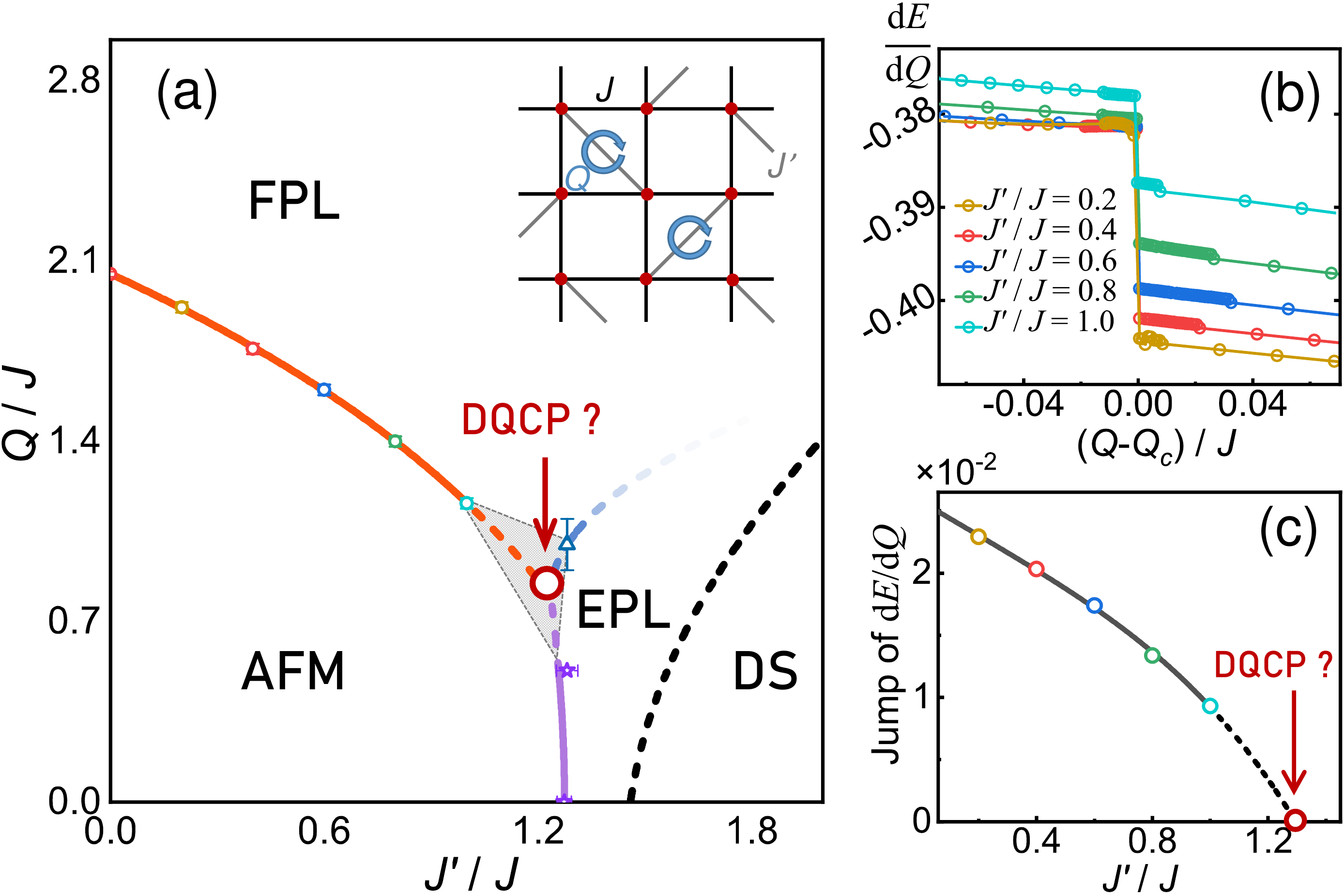} \caption{(a) Ground-state phase diagram of the generalized SS model that preserves the SS lattice symmetry. Besides the DS phase, EPL, FPL, and AFM phases
are stabilized as the ground states. Symbols show the transition points
and the lines are guides to eyes. The triple point of the EPL, FPL,
and AFM phases is proposed as a DQCP (see text). Inset: Sketch of the
generalized SS model with the staggered ring-exchange-like $\hat{H}_{{\rm {Q}}}$
interaction. (b) The first-order derivative of the ground-state energy
$dE/dQ$ with $(Q-Q_{c})/J$, where $Q_{c}$ is the AFM-FPL transition
point. (c) The discontinuity of $dE/dQ$ at $Q=Q_{c}$ versus $J^\prime/J$. The discontinuity vanishes when approaching to the triple point, implying it is a DQCP. \label{fig:4}}
\end{figure}

\textit{Nature of the plaquette ground state.} In the parameter regime $0.68\lesssim J/J^{\prime}\lesssim0.79$,
we are able to stabilize both two plaquette states in the calculation (see Fig.~\ref{fig:1}(d)), and we find the
energy of the EPL state is always lower than that of the FPL. Interestingly,
near the transition point, the energies of these two states become
very close, as shown in Fig.~\ref{fig:3}(a). 
In the large-$D$ limit, the energy difference is 
less than $10^{-4}J$.
Such a small energy difference is about the same order to the one
between the EPL and AFM states near the transition (see inset of Fig.~\ref{fig:2}(a)).
This implies that the FPL state, though never favored as the ground
state of the SS model, emerges as a low-lying competing state near
the plaquette-AFM transition. Actually, besides the FPL, we can also stabilize
other metastable states with competitive energies near the transition.
The emergence of these metastable states suggests enhanced low-energy
fluctuations, and is consistent with the weakly first-order transition
we found.

\textit{Global phase diagram and possible DQCP.} The quasidegeneracy
between the EPL and FPL states prompts that a FPL ground state can
be stabilized nearby if we consider a global phase diagram with an
extra tuning parameter. Such a global phase diagram would help solving
the paradox on the nature of the plaquette ground state between theory and experiments.
A previous theoretical work~\citep{Boos2019} found a quasi-one-dimensional
singlet phase adiabatically connected to the FPL when
the symmetry between the two diagonal $J^{\prime}$ bonds is broken.
But the orthogonal lattice distortion accounting for this symmetry
breaking has not been verified experimentally. Here we adopt a different
strategy, namely, to tune the stability of the EPL and FPL phases
without breaking the SS lattice symmetry. For this purpose, we generalize the SS model
by including a ring-exchange-like interaction.

The model Hamiltonian reads $\hat{H}=\hat{H}_{{\rm {SS}}}+\hat{H}_{{\rm {Q}}}$,
where $\hat{H}_{{\rm {SS}}}$ and $\hat{H}_{{\rm {Q}}}$ are given
in Eqs.~\eqref{eq:SS_Ham} and \eqref{eq:Q_Ham}, respectively. The global phase
diagram in the $J$-$Q$ plane is illustrated in Fig.~\ref{fig:4}(a).
It is shown that the EPL, FPL, and AFM states span a broad regime,
and the ring-exchange term $\hat{H}_{{\rm {Q}}}$ favors the FPL ground
state over the EPL state. We find the FPL-AFM transition is also
weakly first-order, as evidenced by the discontinuity of the ground-state
energy derivative $dE/dQ$ (Fig.~\ref{fig:4}(b) and (c)). Remarkably, going
along the FPL-AFM transition trajectory toward the triple point
among the EPL, FPL, and AFM phases, the discontinuity of the energy derivative
reduces. It is fully suppressed when extrapolating to the the triple
point as shown in Fig.~\ref{fig:4}(c). This implies
that the triple point is a quantum critical point. Note that our model
preserves the SS lattice symmetry and the three phases break distinct
symmetries. We therefore propose this point as a DQCP, because
a continuous transition between any two ordered phases is prohibited
within the LGW paradigm. Again, because the EPL and FPL breaks different
lattice symmetries, their order parameters can be combined to a complex
(two-component) monopole operator~\citep{Lee2019}. Further taking
the three-component spin order parameter of the AFM phase, this infers
an enlarged $SO(5)$ symmetry at the proposed DQCP.

\textit{Discussions and conclusion.} As mentioned in the introduction,
the nature of the transition between the plaquette and AFM phases is under
active debate. To address this issue numerically, it is important
to accurately determine the ground-state energy. In our calculations, the ground state converges quickly with increasing $D$.
Compared to other numerical methods, as shown in Fig.~\ref{fig:3}(a),
the energy of the EPL state at $D=6$ in our calculation is lower
than that of a previous iPEPS work (Ref.~\citep{Corboz2013}) for
$D=7$, and close to the $D=10$ result and a DMRG one~\citep{Lee2019}. For the
AFM state shown in Fig.~\ref{fig:3}(b), the energy at $D=6$ in
our work is lower than those of DMRG~\citep{yang2021quantum} and series expansion results~\citep{Koga2000}.

The discontinuities of $dE/dJ$ and $m_{m}$ unambiguously show a first-order EPL-AFM transition in the SS model, and rule out a continuous transition~\cite{Lee2019} or an intermediate QSL phase~\cite{yang2021quantum}. The extrapolated transition points from different estimates are converged in our calculation. This suggests a QSL regime~\citep{yang2021quantum}, even if exists, should be very narrow in the parameter space. Our results also shed light to the VBS-AFM transitions in the square lattice $J_1$-$J_2$ and checkerboard (CB) models~\citep{Gu_J1J2:2020,Wang_PRL:2018, Brenig_PRB:2002, Fouet_PRB:2003, Berg_PRL:2003, Zou_arXiv:2020}, as both SS and CB models can be realized by depleting the $J_2$ bonds from the $J_1$-$J_2$ model. In the CB model, the full and empty plaquettes have a larger energy difference. We then expect the VBS-AFM transition there to be stronger first-order. In the $J_1$-$J_2$ model, however, the VBS-AFM transition should be weaker, or could even be intervened by an intermediate QSL, because the frustration is not released by depleting the $J_2$ bonds.

In the shaded regime of the global phase diagram in Fig.~\ref{fig:4}, enormous competing states with close energies appear. This prevents us from locating the triple point precisely, and we cannot fully rule out a QSL in this regime. On the other hand, this suggests enhanced fluctuations which usually emerge near a QCP. Details of this phase diagram deserves future investigation.

The purpose of studying the model with the ring-exchange-like interaction in this work is to illustrate the existence of a DQCP by adding a relevant perturbation to the SS model. Certainly a realistic model for $\mathrm{SrCu}_{2}(\mathrm{BO}_{3})_{2}$ may contain more complicated interactions. Our argument for the DQCP is indeed generic: this intriguing physics applies to any perturbation that preserves the SS
lattice symmetry and can tune the ground state between EPL and FPL. Note that a DQCP does not exist in models breaking
the SS lattice symmetry~\citep{Boos2019}, and our proposal obviously
differs from other theoretical suggestions on realizing a DQCP in the
SS model~\citep{Lee2019,yang2021quantum}.

In conclusion, our numerical results unambiguously show a weakly first-order
plaquette-AFM transition in the SS model. Though the ground state favors
the EPL configuration, the FPL is energetically competitive, and can
be stabilized as the ground state under a ring-exchange-like perturbation
that preserves the SS lattice symmetry. Moreover, we provide evidence for a DQCP
at the triple point where the EPL, FPL, and AFM phases meet in the global phase
diagram of the generalized SS model.


\textit{Acknowledgments.} We thank fruitful discussions with Y. Cui, W. Ding, C. Liu, B. Normand, L. Wang, Y. Wang, W. Yu, and H. Zou. This work was supported by the National R\&D Program of China (Grants No. 2017YFA0302900 and No. 2016YFA0300500), the National Natural Science Foundation of China (Grants No. 12174441 and No. 11774420), and the Fundamental Research Funds for the Central Universities and the Research Funds of Renmin University of China (Grants No. 18XNLG24 and No. 20XNLG19).


\clearpage\setcounter{figure}{0} \makeatletter
\global\long\def\thefigure{S\@arabic\c@figure}%
 \onecolumngrid

\setcounter{secnumdepth}{3}


\renewcommand{\thefigure}{S\arabic{figure}}
\renewcommand{\thetable}{S\Roman{table}}
\renewcommand{\theequation}{S\arabic{equation}}
\renewcommand{\thesection}{S \Roman{section}}
\renewcommand{\bibnumfmt}[1]{[S#1]}
\renewcommand{\citenumfont}[1]{S#1}

\section*{SUPPLEMENTAL MATERIAL -- First-order transition between the plaquette valence bond solid and
antiferromagnetic phases of the Shastry-Sutherland model}

\subsection{Variational Optimization for the 16-PESS tensor network states}

To find the tensor-network-state
representation of a target state, many optimization methods, including
time evolution with simple update (SU) [30,31,35],
cluster update (CU)[39],
full update(FU)[37,38], 
and traditional variational optimization [38,41],
have been developed. In this work, we adopt an advanced
variational optimization of the tensor network states with the projected entangled simplex state (PESS) 
construction. This is inspired
by the differentiable programming [32-34], 
which has shown great potential in the study of classical systems
and quantum many-body systems. Here we construct the 16-PESS tensor
network states on a square lattice as shown in Fig. 1(b) of the main
text. We choose a 4-site cluster as a unit cell expressed by a $U$
tensor. Every four adjacent units are connected by an entangled simplex
tensor $S$. The simplex tensors $S$ are used to describe the entanglement
of these clusters. The $U$ and $S$ tensors are all placed at plaquettes
with the $J^{\prime}$ coupling.

The optimization procedure starts from a PESS tensor network state, which can be
obtained from a random initilization or a state after time evolution (combined with SU, CU, and FU).
The variational process further optimizes
the expectation value of the energy associated with the PESS state
iteratively, until the energy is converged. The state with the lowest
energy will be eventually selected as the candidate ground state of
the system for a given auxiliary bond dimension $D$. Deep inside
a phase, the nature of the ground state and the ground-state energy
is not sensitive to the way of initialization. However, when the system
is close to a phase transition, the converged state in usual optimization
process may heavily depend on the initial states, as being trapped
in local minimum is a very common problem. To relieve this issue,
careful variational treatment is very important.

\begin{figure}[h]
\includegraphics[width=0.72\linewidth]{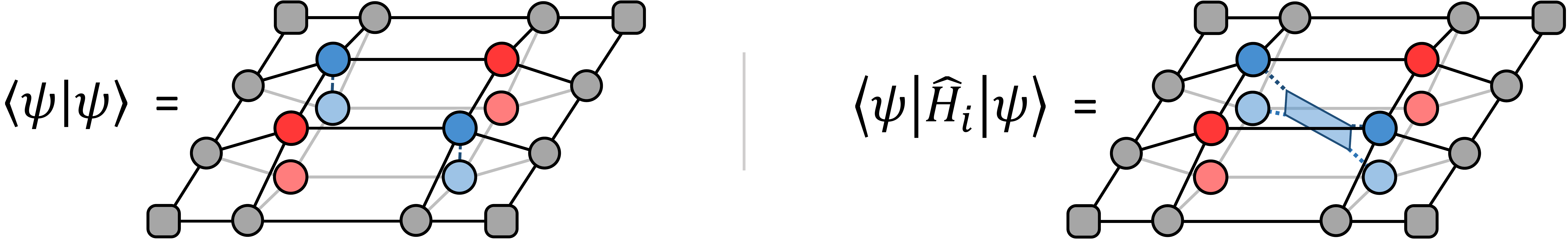} \caption{Tensor-diagram representation of the norm of the variational state
${\langle\psi|\psi\rangle}$ and the expectation value of the local
energy $E_{i}={\langle\psi|\hat{H_{i}}|\psi\rangle}/{\langle\psi|\psi\rangle}$.
The blue and red dots refer to the $U$ and $S$ tensors in the 16-PESS
construction, respectively, and the grey dots denote the environment
tensors constructed by CTMRG. The blue rectangle sandwiched between
the two layers of network in the right panel refers to the local Hamiltonian
operator $H_{i}$. \label{fig:S1}}
\end{figure}

The basic idea of our variational process is to minimize the ground-state
energy density $E_{i}={\langle\psi|\hat{H_{i}}|\psi\rangle}/{\langle\psi|\psi\rangle}$
globally. Within the 16-PESS tensor network construction, $E_{i}$
and the norm of the variational state $\langle\psi|\psi\rangle$ can
be effectively evaluated, as illustrated in Fig.~\ref{fig:S1}. The
total ground-state energy $E$ is then obtained by summing over the
local energies. For the specific details of the optimization, we use
the corner transfer matrix renormalization group (CTMRG) [38] 
method to contract the infinite network and get the approximate environment
of the local tensors. After that, we use the quasi-Newton L-BFGS algorithm
to search for the minimum of energy expectation value, in which an
automatic gradient method supported by Zygote [40] 
is applied.
This automatic differentiation program can compute $\partial E_{i}/\partial U(S)$
effectively. This process involves nonlinear optimization and it is
the key point for a global minimization.

\begin{figure*}[h!]
\includegraphics[width=0.82\linewidth]{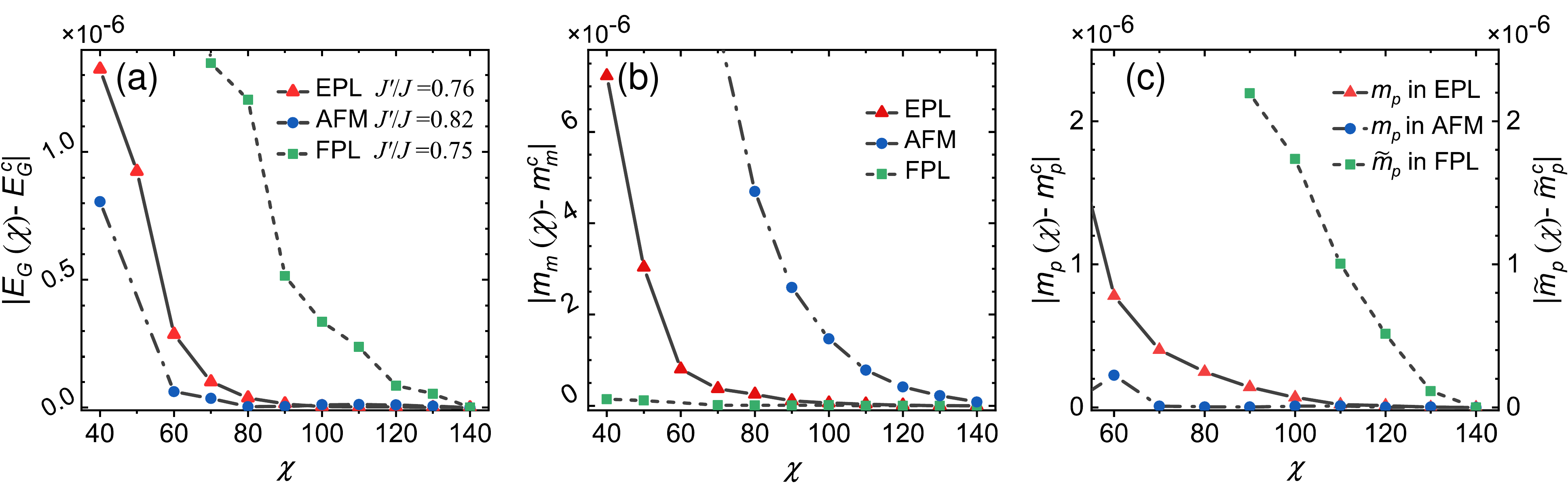} \caption{Convergence of expectation values versus $\chi$ with $D=5$.  The corner mark $c$ here denotes the final (convergent) value (with the largest $\chi$ we employed). (a): Convergence of
the energy as a function of $\chi$. (b): Convergence of the antiferromagnetic order parameter
$m_{m}$ as a function of $\chi$. (c): Convergence
of the plaquette order parameters $m_{p}$($\widetilde{m}_{P}$) as
a function of $\chi$. Here $\widetilde{m}_{P}$ is the order parameter
of the FPL state, defined as $\widetilde{m}_{P}=\sum\limits _{\left\langle ij\right\rangle \in\boxslash}\langle\mathbf{S}_{i}\cdot\mathbf{S}_{j}\rangle-\sum\limits _{\left\langle ij\right\rangle \in\boxbslash}\langle\mathbf{S}_{i}\cdot\mathbf{S}_{j}\rangle$
and $m_{p}$ is the order parameter of the EPL state, defined in Eq.
(4) of the main text. \label{fig:S2}}
\end{figure*}

When the optimization is converged, we further use the CTMRG to contract
the infinite network and calculate the expectation value of energy.
To obtain a more reliable extrapolation and comparison, we push the
CTMRG truncation dimension $\chi$ to a sufficiently large value.
In the practical calculation, we use $\chi$ as large as $160$ for convergence.
As demonstrated in Fig.~\ref{fig:S2},
all the quantities are converged 
within $10^{-7}$.

\subsection{Fitting And Extrapolation}

We can estimate the transition point from the discontinuity of the
order parameters as shown in Fig. 2 in the main text. To further confirm
the result, we can use the level-crossing method described as follows,
to obtain a more accurate result.

\begin{figure*}[h!]
\includegraphics[width=0.8\linewidth]{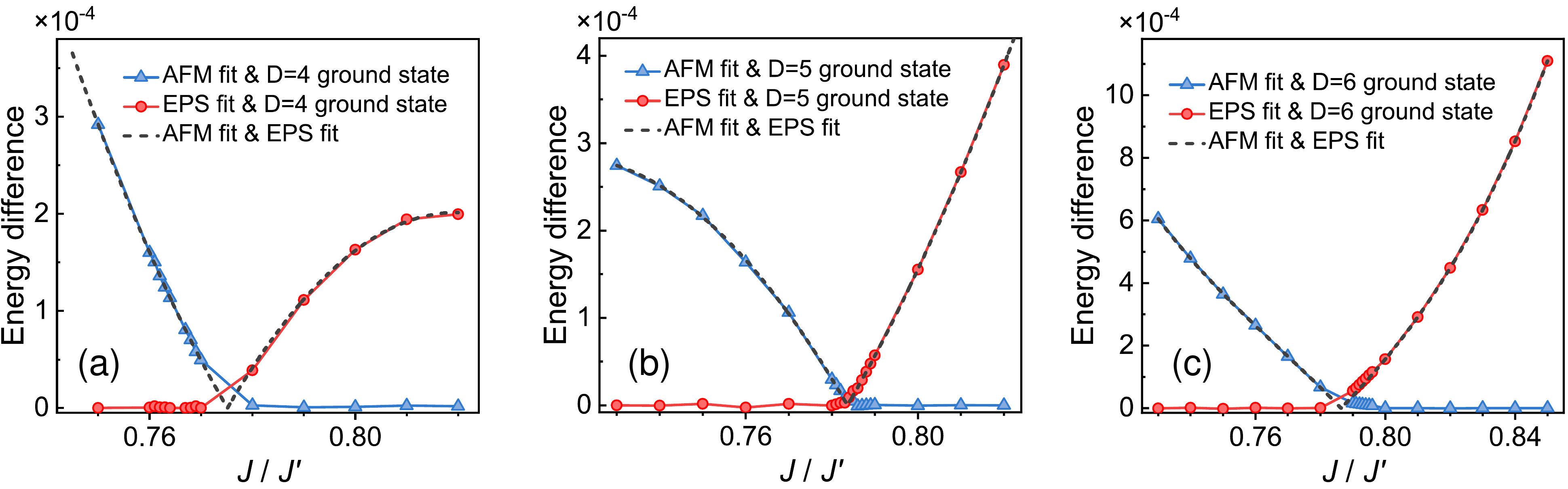} \caption{The energy differences $\delta E_{A}$ and $\delta E_{P}$ versus
$J/J^{\prime}$ at $D=$ 4, 5, 6 for (a), (b), (c) respectively. \label{fig:S3}}
\end{figure*}

For each $D$, we fit the energies obtained deep in each phase to
a cubic polynomial function. Define the energy difference $\delta E_{P(A)}=E-E_{P(A)}$,
where $E$ is the calculated energy, and $E_{P(A)}$ is the fitted
energy in the PL (AFM) state. The transition point can then be determined
as the crossing point of $\delta E_{P}$ and $\delta E_{A}$ for this
transition (see also main text). The crossing points for $D=$4, 5,
6 are shown in Fig. \ref{fig:S3}, respectively. These data are then
used for extrapolating the transition point in the large-$D$ limit,
as shown in Fig. 2(f) of the main text. The transition point for infinite
$D$ value is then estimated to be $J/J^{\prime}\approx0.79$ from
both exponential and power-law fittings.

\end{document}